\documentclass[aps,prd,twocolumn, superscriptaddress,nofootinbib]{revtex4}
\usepackage{amsfonts}
\usepackage{mathtools}
\usepackage{hyperref}
\usepackage{graphicx}
\usepackage{dcolumn}
\usepackage{bm}
\usepackage{tensor}

\begin{document}


\title{Spacetime Curvature in terms of Scalar Field Propagators}

\author{Mehdi Saravani}
\email{msaravani@perimeterinstitute.ca}
\affiliation{Perimeter Institute for Theoretical Physics, 31 Caroline St. N., Waterloo, ON, N2L 2Y5, Canada}
\affiliation{Department of Physics and Astronomy, University of Waterloo, Waterloo, ON, N2L 3G1, Canada}

\author{Siavash Aslanbeigi}%
\email{saslanbeigi@perimeterinstitute.ca}
\affiliation{Perimeter Institute for Theoretical Physics, 31 Caroline St. N., Waterloo, ON, N2L 2Y5, Canada}
\affiliation{Department of Physics and Astronomy, University of Waterloo, Waterloo, ON, N2L 3G1, Canada}

\author{Achim Kempf}
\email{akempf@uwaterloo.ca}
\affiliation{Department of Applied Mathematics, University of Waterloo, Waterloo, Ontario, N2L 3G1, Canada}
\affiliation{Institute for Quantum Computing, University of Waterloo, Waterloo, Ontario, N2L 3G1, Canada}
\affiliation{Department of Physics and Astronomy, University of Waterloo, Waterloo, ON, N2L 3G1, Canada}
\affiliation{Perimeter Institute for Theoretical Physics, 31 Caroline St. N., Waterloo, ON, N2L 2Y5, Canada}

\date{\today}

\begin{abstract}
We show how quantum fields can be used to measure the curvature of spacetime. In particular, we find that knowledge of the imprint that spacetime curvature leaves in the correlators of quantum fields suffices, in principle, to reconstruct the metric. We then consider the possibility that the quantum fields obey a natural ultraviolet cutoff, for example, at the Planck scale. We investigate how such a cutoff limits the spatial resolution with which curvature can be deduced from the properties of quantum fields. We find that the metric deduced from the quantum correlator exhibits a peculiar scaling behavior as the scale of the natural UV cutoff is approached.  
\end{abstract}

\maketitle


\section{Introduction}


In general relativity, spacetime measurements are traditionally based on the use of some form of standard rods and clocks. At sub-atomic scales, there are of course no rods or clocks in Einstein's sense and the only available tools then are quantum fields. We will, therefore, here address the question how the curvature of a classical spacetime can be expressed solely through in-principle measurable properties of quantum fields. The ability to express the curvature of a classical spacetime entirely in terms of quantized degrees of freedom of fields could become a useful tool in the quest to then also quantize the spacetime curvature itself, see e.g., \cite{Gibbons}-\cite{Loll}. We will also take into account that quantum fields are likely subject to a natural ultraviolet cutoff at the Planck scale (for a review, see, e.g. \cite{hossenfelder}). We will study how such a cutoff limits the spatial resolution with which the spacetime metric can be deduced from in-principle measurable properties of a quantum field. 

We begin by recalling that the curvature of a classical spacetime influences not only matter and radiation but also the vacuum, \cite{BirrellDavies}. This is because curvature influences wave operators such as the d'Alembertian and curvature therefore also impacts the normal mode decomposition of quantum fields. This means that curvature affects the vacuum state of quantum fields, affecting for example, the vacuum entanglement and the correspondingly correlated quantum fluctuations between different locations in spacetime \cite{valentini,sorkin,srednicki,menicucci,ak-entharvesting,ak-cliche2}. 

The question that we address here is whether knowledge solely of this imprint that curvature leaves on the quantum vacuum is sufficient to be able to deduce the curvature of the spacetime. This is nontrivial, considering that, for example, knowledge of only the energy-momentum tensor of quantum fields would be insufficient because the energy momentum tensor determines only the Ricci component but not the Weyl component of the curvature. On the other hand, it is known that spacetime curvature affects interacting quantum fields to the extent that counter terms are induced that include the Einstein action in the leading orders. Spacetime curvature therefore affects quantum fields sufficiently to induce Einsteinian dynamics \cite{Sakharov,Visser}. 

Our first finding here is that the impact that spacetime curvature has on the statistics of the quantum fluctuations of a scalar field is in fact complete. Concretely, the knowledge of even just the spacetime-dependent propagator of a free scalar field on a curved spacetime suffices to calculate the metric on the spacetime and therefore to obtain the spacetime curvature. 
The propagator is  part of the Feynman rules and, in principle, in a curved spacetime, the correspondingly spacetime-dependent propagator can be inferred with suitable particle physics experiments. This then replaces standard rods and clocks. 

For intuition, let us consider that propagators are correlators. This means that by considering a propagator we are considering the impact that curvature has on the spatial and temporal correlations of vacuum fluctuations of quantum fields. Why then should the correlator yield metric information? Intuitively, the reason is that the strength of the correlations of spatially and temporally separated quantum vacuum fluctuations provides a measure of spacetime distance, and knowing distances is to know the metric, as has been argued in \cite{ak-njp}.

For an alternative perspective, let us recall that knowledge of the light cones of a spacetime allows one to deduce the spacetime metric up to a local conformal factor \cite{HawkingEllis}. In effect, our result is that a scalar quantum field's propagator does not only indicate the light cones, but also the local conformal factor.  

Having established a straightforward method to extract the metric from a propagator, we then consider the case where the quantum field is subject to a natural ultraviolet cutoff. In this case, it should not be possible to use quantum fields to probe the curvature at length scales that are smaller than the cutoff scale. To this end, we use a simple model for the natural ultraviolet cutoff, namely a hard cutoff within the framework of Euclidean-signature quantum field theory. We examine how the metric that is deduced from quantum field correlators behaves as the natural UV cutoff scale is approached. We find characteristic oscillations that are generally unobservable because they are washed out by the cutoff. However, through the fluctuation amplifying effects of cosmic inflation, see, e.g., \cite{Liddle}, such oscillations in the metric may conceivably have left a signature in the cosmic microwave background. 

%
%
%
%
%
%

\section{Deducing the metric from the propagator}

\subsection{Flat space}\label{Flat space}
We begin with the simple case of  quantum field theory in flat Euclidean space. The aim is to determine if the metric tensor can be reconstructed from the correlator of a scalar quantum field. To this end, we recall that in $D$-dimensional Euclidean space, the massive Green's function satisfies
\begin{equation}
(\nabla_x^2-m^2)G(x,y)=-\delta^{(D)}(x-y),
\end{equation}
and that $G(x,y)$ is given explicitly by 
\begin{equation}
G(x,y)=\frac{(2\pi)^{-\frac{D}{2}}}{r_{xy}^{D-2}}(mr_{xy})^{\frac{D}{2}-1}K_{\frac{D}{2}-1}(mr_{xy}),
\label{eucmassiveG}
\end{equation}
where $K_{\nu}(x)$ is the modified Bessel function of the second kind, and 
\begin{equation}
r_{xy}^2=|x-y|^2=\sum_{i=1}^{D}(x^i-y^i)^2.
\end{equation}
In the massless limit, $G(x,y)$ takes the form
\begin{equation}
G(x,y)\xrightarrow{mr_{xy}\to0}G_0(x,y)=
\frac{\Gamma(D/2-1)}{4\pi^{D/2}r_{xy}^{D-2}}.
\label{masslesslim}
\end{equation}
Since $G_0(x,y)$ depends quite simply on the distance, $r_{xy}$, we can easily use $G_0(x,y)$ to reconstruct the flat metric:
\begin{align}
\delta_{ij}&=-\frac{1}{2}\frac{\partial}{\partial x^i}\frac{\partial}{\partial y^j}
r_{xy}^2\notag\\
&=-\frac{1}{2}\left[\frac{\Gamma(D/2-1)}{4\pi^{D/2}}\right]^{\frac{2}{D-2}}
\frac{\partial}{\partial x^i}\frac{\partial}{\partial y^j}
\left(G_0(x,y)^{\frac{2}{2-D}}\right).
\end{align}
Let us now ask if a massive field's Green's function can also directly be used 
to recover the metric. Intuitively, one expects this to be true because mass mostly affects the infrared
and should matter little when $|x-y|\ll m^{-1}$. Indeed, one can verify that, in the ultraviolet limit, $x\to y$:
\begin{equation}
\delta_{ij}=-\frac{1}{2}\left[\frac{\Gamma(D/2-1)}{4\pi^{D/2}}\right]^{\frac{2}{D-2}}
\lim_{x\to y}
\frac{\partial}{\partial x^i}\frac{\partial}{\partial y^j}
\left(G(x,y)^{\frac{2}{2-D}}\right).
\label{eucGtog}
\end{equation}
Let us consider, for example, the case $D=3$ in which \eqref{eucmassiveG}
simplifies to
\begin{equation}
G(x,y)=\frac{e^{-mr_{xy}}}{4\pi r_{xy}}.
\label{Euc3G0}
\end{equation}
It can be verified in this case that the RHS of
\eqref{eucGtog} (without the limit) is given by:
\begin{align}
&-\frac{1}{2}\left[\frac{\Gamma(1/2)}{4\pi^{3/2}}\right]^{2}
\frac{\partial}{\partial x^i}\frac{\partial}{\partial y^j}
\left(G(x,y)^{-2}\right)\notag\\
&=e^{2mr}
\left[(1+mr)\delta_{ij}
+\frac{m}{r}(3+mr)\delta_{ik}\delta_{jl}
(x^k-y^k)(x^l-y^l)\right].
\end{align}
Taking the $x\to y$ limit we find $\delta_{ij}$.
In Appendix \ref{app:GtogDEuc}, we show  
that \eqref{eucGtog} is true also for $D\ge4$.


\subsection{Curved space} 
Our aim now is to express the metric in terms of the correlator of quantum fluctuations of a scalar field in curved manifolds. Then,  
the Green's function satisfies the equation
\begin{equation}
(\Delta_x-m^2)G(x,y)=-\frac{\delta^{(D)}(x-y)}{\sqrt{g(x)}},
\label{GFcurvedDE}
\end{equation}
where $\Delta_x=\frac{1}{\sqrt{g(x)}}\partial_{x^i}\left(\sqrt{g(x)}g^{ij}(x)\partial_{x^j}\right)$ is the Laplace-Beltrami operator.
As we will show, \eqref{eucGtog} therefore straightforwardly generalizes to curved manifolds:
\begin{equation}
{
g_{ij}(y)=-\frac{1}{2}\left[\frac{\Gamma(D/2-1)}{4\pi^{D/2}}\right]^{\frac{2}{D-2}}
\lim_{x\to y}
\frac{\partial}{\partial x^i}\frac{\partial}{\partial y^j}
\left(G(x,y)^{\frac{2}{2-D}}\right)
}
\label{GtogCurved}
\end{equation}
First, let us confirm that \eqref{GtogCurved} does not depend on the coordinate system, i.e., that it represents a covariant way to express the metric in quantum terms. 
To this end, consider two coordinate systems $x$ and $\tilde x$ and the Green's function in each coordinate $G(x,y)$ and $\tilde G(\tilde x,\tilde y)$, respectively. Since $G$ is a bi-scalar, $\tilde G(\tilde x,\tilde y)=G(x,y)$:
\begin{align}
&\tilde g_{ij}(\tilde x)=-\frac{1}{2}\left[\frac{\Gamma(D/2-1)}{4\pi^{D/2}}\right]^{\frac{2}{D-2}}
\lim_{\tilde y \to \tilde x}
\frac{\partial}{\partial \tilde x^i}\frac{\partial}{\partial \tilde y^j}
\left(\tilde G(\tilde x,\tilde y)^{\frac{2}{2-D}}\right)\notag\\
&=-\frac{1}{2}\left[\frac{\Gamma(D/2-1)}{4\pi^{D/2}}\right]^{\frac{2}{D-2}}
\lim_{\tilde y \to \tilde x}
\frac{\partial}{\partial \tilde x^i}\frac{\partial}{\partial \tilde y^j}
\left(G( x, y)^{\frac{2}{2-D}}\right)\notag\\
&=-\frac{1}{2}\left[\frac{\Gamma(D/2-1)}{4\pi^{D/2}}\right]^{\frac{2}{D-2}}
\lim_{y \to x}
\frac{\partial x^k}{\partial \tilde x^i}\frac{\partial y^l}{\partial \tilde y^j}\frac{\partial}{\partial x^k}\frac{\partial}{\partial y^l}
\left(G( x, y)^{\frac{2}{2-D}}\right)\notag\\
&=\frac{\partial x^k}{\partial \tilde x^i}\frac{\partial x^l}{\partial \tilde x^j}g_{kl}(x).
\end{align}
Now in order to verify \eqref{GtogCurved} it is instructive to work out a special case in detail, such as the case of  the $D$-sphere where the Green's function is explicitly known. In Appendix \ref{app:GtogDSphere} we show that \eqref{GtogCurved} holds in this case. 

In order to prove that \eqref{GtogCurved} holds on all Riemannian and also on all pseudo-Riemannian manifolds, it is fortunately not necessary to know the Green's function explicitly. Due to the presence of the limit $x\to y$
in \eqref{GtogCurved}, it suffices to know
the behaviour of $G(x,y)$ when $x$ and $y$
are arbitrarily close. In this regime,
$G(x,y)$ takes its flat space form plus
corrections which arise due to curvature and which are benign in the limit $x\to y$.
Based on this idea, we give the detailed proof of \eqref{GtogCurved} for all  (pseudo-) Riemannian manifolds in Appendix \ref{app:GtogCurved}.

\section{Introduction of a covariant UV cutoff}
Eq.\eqref{GtogCurved} shows how the metric and therefore the curvature can be reconstructed from its effect on quantum fields. What, however, if the quantum fields are subject to a natural ultraviolet cut-off, in which case the matter degrees of freedom cannot be used to resolve any structure that is smaller than, for example, a Planck length? How does the metric that one reconstructs from the matter degrees of freedom behave then, in particular, towards the ultraviolet? 

\subsection{A model for a covariant UV cutoff}
Due to lack of experimental evidence, it is not known how spacetime behaves close to the Planck scale. 
It has been argued, for example, that spacetime is discrete at that scale, see e.g.,  \cite{Rovelli}. Technically this could regulate ultraviolet divergences and it is consistent with the fact that quantization literally and quite often concretely means discretization. But it has also been argued that, as general relativity seems to indicate, spacetime should remain continuous at all scales. This would not help with ultraviolet divergencies but it would preserve symmetries that lattices break. It would also avoid, for example, potential problems of non-adiabaticity associated with discrete point production during cosmic expansion \cite{jacobson}. 

But there is also the possibility that spacetime is simultaneously both continuous and discrete, namely in the same mathematical way that information can be \cite{ak-shannon2000,ak-shannon2004,ak-shannon2009,ak-njp}. This could combine the advantages of both pictures. To see this, let us recall  Shannon's sampling theorem from information theory. The theorem establishes the equivalence between continuous and discrete representations of information and it is in ubiquitous use in digital signal processing and communication engineering. Assume that a signal, $f$, representing continuous information, is bandlimited, i.e., it consist of frequencies only within a finite frequency range $(-\Omega,\Omega)$:
\begin{equation}
f(t) = \frac{1}{\sqrt{2\pi}}\int_{-\Omega}^\Omega \tilde{f}(\omega) ~e^{i\omega t} ~d\omega.
\end{equation}
Shannon's theorem holds that it suffices to record the signal at a discrete set of times $t_n$ with spacing $t_{n+1}-t_n=\pi/\Omega$ to capture the signal completely. Namely, $f(t)$ can actually be perfectly reconstructed at all times $t$ from the discrete samples $\{f(t_n)\}$ of the signal:
\begin{equation}
f(t) = \sum_n f(t_n) \frac{\sin((t-t_n)\Omega)}{(t-t_n)\Omega}.
\end{equation}
In fact, the signal can be perfectly reconstructed from any set of samples, even non-equidistantly chosen samples, if the average density (technically, the Beurling density) of samples is at least $\Omega/\pi$ (although non-equidistant sampling comes at the cost of an increased sensitivity of the reconstruction to inaccuracies in the recording of the samples). 

Physical fields could be spatially bandlimited in the same way, namely if there exists a suitable natural ultraviolet cutoff in nature \cite{ak-shannon2000,ak-shannon2004,ak-shannon2009}. This is a simple model of how quantum fields may behave towards the Planck scale. But it is also the second-quantized manifestation of what is in first quantization the minimum length uncertainty principle which has long been suggested to arise from various approaches to quantum gravity, see e.g., \cite{Garay}, including string theory, see e.g. \cite{witten} and quantum groups \cite{ak-qgroupsucr,ak-kmm,ak-shannon2000} which arise in noncommutative geometry \cite{Majid}. 

If we assume this type of natural ultraviolet cutoff, physical fields are defined on a continuous spacetime, as usual, but it suffices to know a field on a sufficiently dense lattice to be able to reconstruct the field everywhere. Crucially, since any sufficiently densely-spaced lattice can be chosen, the symmetries of the continuous spacetime are preserved. 
Physical fields in euclidean-signature spaces are then considered covariantly bandlimited if they are in the span of those eigenfunctions of the Laplacian whose eigenvalues (playing the role of squared spatial frequencies) are below a cut-off value of $\Lambda$, where $\Lambda$ may be, for example, the square of the Planck momentum. 

Recently, it has been shown how the entanglement entropy of quantum fields can be calculated within this framework, and that it exhibits the expected scaling laws \cite{ak-pye-donnelly}. 

Here, let us consider the impact of this type of natural ultraviolet cutoff on the extend to which the metric deduced from the propagator can be spatially resolved. 
Concretely, consider the eigenvalues ($\lambda_n\in\mathbb{R}$) and eigenfunctions of the Laplacian (with or without mass term) on
an arbitrarily curved Riemannian manifold $M$
\begin{equation}
(\Delta - m^2) f_n(x)=-\lambda_n^2f_n(x), \qquad
\langle f_n,f_m \rangle = \delta_{nm},
\end{equation}
where $\langle \cdot,\cdot \rangle$ is the $L^2$ inner product: $
\langle f,h \rangle=\int_Mf(x)h(x)\sqrt{g}d^Dx$.
For simplicity, namely so that the self-adjoint extension of the Laplacian is unique and so that the eigenvalues are discrete, we are here assuming that the Riemannian manifold is compact without boundaries. 
The associated Green's function, which satisfies \eqref{GFcurvedDE},
can be written expanded in the eigenbasis of the Laplacian:
\begin{equation}
G(x,y)=\sum_n\frac{1}{\lambda_n^2}f_n(x)\bar{f}_n(y).
\end{equation}
By implementing the ultraviolet cutoff, we now obtain the \textit{bandlimited} Green's function  $G_{\Lambda}(x,y)$: 
\begin{equation}
G_{\Lambda}(x,y)=\sum_n^{\lambda_n<\Lambda}\frac{1}{\lambda_n^2}f_n(x)\bar{f}_n(y).
\end{equation}
Our method of above for expressing the metric in terms of the quantum correlator can now be applied to this UV cutoff Green's function, i.e., we apply \eqref{GtogCurved} to $G_\Lambda$ to obtain a modified metric $g^\Lambda$: 
\begin{equation}\label{BLmetric}
g^\Lambda_{ij}(y)\equiv-\frac{1}{2}\left[\frac{\Gamma(D/2-1)}{4\pi^{D/2}}\right]^{\frac{2}{D-2}}
\lim_{x\to y}
\frac{\partial}{\partial x^i}\frac{\partial}{\partial y^j}
G_{\Lambda}(x,y)^{\frac{2}{2-D}}.
\end{equation}
We can now address the question that we set out to answer, namely the question how the UV cutoff impacts the expression of the metric in terms of the correlator. 

\subsection{Impact of the covariant UV cutoff on the reconstructed metric: flat space}
First let us consider again the case of a massless scalar field in flat $\mathbb{R}^D$. To this end, we start with the Green's function with the ultraviolet cutoff implemented:
\begin{equation}\label{BLGreenflat}
G_{\Lambda}(x,y)=\int_{|p|<\Lambda}\frac{d^Dp}{(2\pi)^D}\frac{1}{p^2}e^{ip\cdot(x-y)}.
\end{equation}
Substituting \eqref{BLGreenflat} in \eqref{BLmetric}, we obtain the following metric
\begin{equation}\label{BLmetricflat}
g_{ij}^{\Lambda}=\frac{4}{D^2}\Gamma[D/2]^{\frac{4}{D-2}}\delta_{ij}.
\end{equation}
The details of the calculation are in Appendix \ref{app:BLflat}. 
This result shows that one recovers the flat metric, $\delta_{ij}$, but only up to a constant prefactor $\nu(D)=\frac{4}{D^2}\Gamma[D/2]^{\frac{4}{D-2}}$. Notice that the prefactor $\nu(D)$ is independent of the UV cutoff, i.e., it persists even when the UV cutoff, $\Lambda$, is sent to infinity, $\Lambda\rightarrow \infty$. 
Interestingly, this means that, in  \eqref{BLmetric}, the UV limit $\Lambda \rightarrow \infty$ does not commute with the UV limit $x\rightarrow y$. This is made possible by the fact that the Green's function is UV divergent as $x\rightarrow y$ without the cutoff but becomes a regular function in $x$ and $y$ with the UV cutoff implemented. 

Given that the prefactor $\nu(D)$ 
is a UV phenomenon, we expect that it also appears on all curved spacetimes, so long as there is no significant curvature close to the UV cutoff scale. We present concrete evidence 
for this expectation in Appendix \ref{app:BLS3}, where we show that applying \eqref{BLmetric} to the
3-Sphere yields the correct metric with the same prefactor $\nu(D)$, which in three dimensions reads $\nu=\frac{\pi^2}{36}$. Because of this feature, we shall refer to $\nu(D)$ as the \textit{universal prefactor}.

Universality of $\nu(D)$ suggests that \eqref{BLmetric}, once
corrected by the overall scaling $\nu(D)$, yields a methodology for
``smoothening out" a Riemannian metric on the length scale $1/\Lambda$.
This may prove to be useful as a mathematical tool in quantum gravity,
where integrating out the metric degrees of freedom is of interest.
Note that smoothening out the metric on a given length scale is 
non-trivial because the metric is what \textit{defines} length scales.
Here we arrived at a smoothening method for the metric by using two key properties of the Green's function: it encodes distances and it is straightforward to implement the UV cutoff in the Green's function. In Section \ref{FRW}, we apply our methodology to explicitly demonstrate how a wiggly manifold's metric is indeed smoothened out by adopting the metric deduced from the propagator in which the UV cutoff has been implemented.

Before moving on to Section \ref{FRW}, let us discuss the 
possible physical origin and consequences of the universal prefactor.

\subsection{Oscillations in the reconstructed metric without performing the coincidenc limit}
Let us recall that our method for expressing the metric in terms of the two-point function of a scalar quantum field works accurately when there is no ultraviolet cutoff. However, we also found that in the presence of a natural UV cutoff, our method, \eqref{BLmetric}, recovers the metric up to a prefactor. 

In fact, as we will now show, our method recovers the metric correctly, i.e., without any need for a corrective prefactor, once we properly take into account all implications of the presence of a natural UV cutoff. Namely, if there is a natural UV cutoff then distances smaller than the cutoff scale cannot be resolved. This means that in \eqref{BLmetric} the limit $x\rightarrow y$ should not be taken, given that it has no operational meaning in terms of measurable quantities. 

Let us, therefore, consider the right hand side of \eqref{BLmetric} but without taking the limit. Instead, let us view the right hand side as a function of the distance between $x$ and $y$. 

In addition, to be fully consistent with the presence of the natural ultraviolet cutoff, we should also not take the Newton Leibniz limit that is implicit in the taking of the two derivatives in \eqref{BLmetric}. It will be instructive, however, to first study the case where the Newton Leibniz limits are taken. 

We will first consider the case of 3-dimensional euclidean space. This case is representative for all those situations in which  curvature is significantly present only at length scales that are significantly larger than the length scale of the natural ultraviolet cutoff. 

To this end, we recall that in the case of 3-dimensional flat space, the bandlimited Green's function is given by
\begin{equation}
G(x,y)=\frac{1}{2\pi^2}\frac{Si(\Lambda r_{xy})}{r_{xy}}, \label{tr1}
\end{equation}
where
$\Lambda$ is the cutoff, $r_{xy}=|x-y|$ and $Si(x)=\int_0^x dt~\frac{\sin(t)}{t}$.

\begin{figure}
\centering
\includegraphics[scale=0.45]{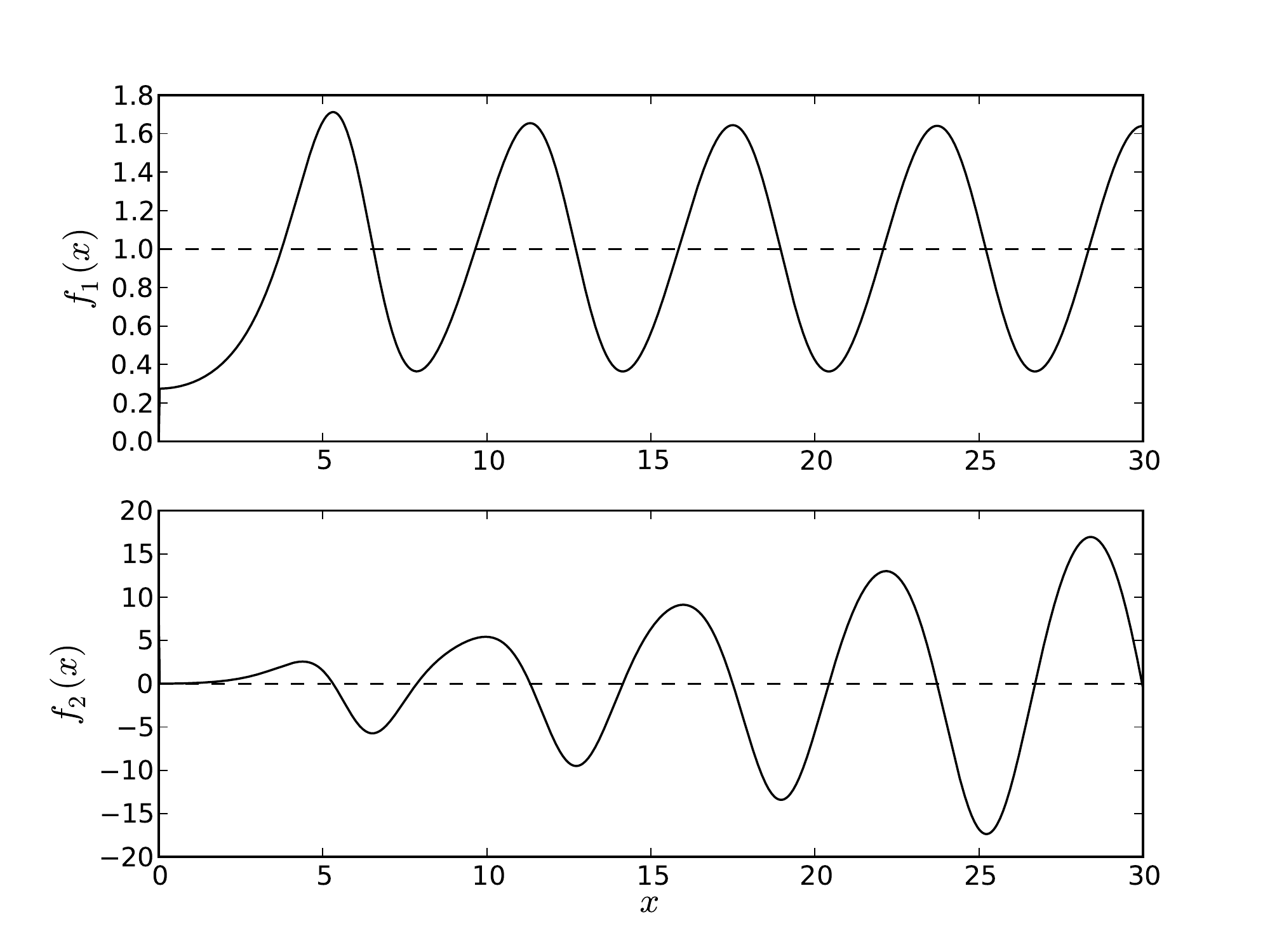}
\caption{Behaviour of $f_1(x)$ and $f_2(x)$, which are defined
         in \eqref{f1}--\eqref{f2}.}
\label{fig:osc}
\end{figure}

Our Green's function-to-metric method, now without performing the coincidence limit, but still performing the Newton Leibniz limits that are part of the derivatives, yields
\begin{align}
g_{\alpha \beta}(x,y)&=\delta_{\alpha\beta}f_1(\Lambda r_{xy})
+\frac{(x_{\alpha}-y_{\alpha})(x_{\beta}-y_{\beta})}{r_{xy}^2}f_2(\Lambda r_{xy}),\\
f_1(x)&=\frac{\pi^2}{4}\left(\frac{1}{Si^2(x)}-\frac{\sin(x)}{Si^3(x)}\right),\label{f1}\\
f_2(x)&=\frac{\pi^2}{4}\left(3\frac{\sin^2(x)}{Si^4(x)}-2\frac{\sin(x)}{Si^3(x)}-\frac{\Lambda r\cos(x)}{Si^3(x)}\right). \label{f2}
\end{align}
If we perform the coincidence limit, we obtain the expected prefactor $\nu$:
\begin{equation}
g_{\alpha \beta}(x)=\lim_{y\rightarrow x}g_{\alpha \beta}(x,y)=\frac{\pi^2}{36}\delta_{\alpha \beta}.
\end{equation}
However, as is shown in FIG. \ref{fig:osc}, the `metric' $g_{\alpha\beta}(x,y)$ oscillates as $y$ approaches $x$ from larger distances. 
The oscillations have wavelengths at the cutoff scale. This could only happen because we did perform the Newton Leibniz limit inside the derivatives in \eqref{BLmetric}, as if there were no ultraviolet cutoff.

In fact, of course, these oscillations cannot actually be resolved, due to the fact that differences of distance as small as the Planck scale cannot be resolved in the presence of the natural UV cutoff. 
With the precision that is accessible, these oscillations are washed out and only their average value matters. That average value is unity, which means that the metric is in fact recovered from the Greens function with a prefactor of one via \eqref{BLmetric}, when working with only the precision that is available in the presence of the ultraviolet cutoff.

Now, interestingly, we arrived at this conclusion under the assumption that there is curvature only at length scales that are significantly larger than the length scale of the natural ultraviolet cutoff. However, for example, in inflationary cosmology, the Hubble radius during inflation is thought to have been only about five or six orders of magnitude larger than the Planck length. It is conceivable, therefore, that Planck scale physics could impact to some extent the predictions for the CMB. Inflation may have acted as a magnifying glass to make the above-discussed oscillations visible in the CMB. 

\subsection{Example of the curvature scale reaching the cut-off scale}
\label{FRW}
Let us investigate the implications for constructing the metric from the Green's function when there exists significant curvature down to scales close to the cutoff scale. How is this curvature smoothed out? We choose a simple example in three dimensions:
\begin{equation}
ds^2=a^2(\eta)(d\eta^2+dx^idx^i),
\end{equation}
where $a(\eta)=1+\epsilon(\eta)$ and $\epsilon(\eta)\ll1$. Let's call $\eta$ 'time' coordinate just to distinguish it from $x^i$ coordinates. 
We use Greek letters for all coordinates and Latin ones only for 'spatial' coordiantes. The aim is to investigate the effect of the UV cutoff when the UV cutoff and the curvature scales are not well separated. We further assume that the perturbation $\epsilon$ exists only in a finite interval of $\eta$ and that
\begin{equation}\label{ep0}
\int \epsilon(\eta) d\eta=0.
\end{equation}
The Laplace operator is given by
\begin{equation}
\Delta=a^{-2}\partial_{\eta}^2+\frac{a'}{a^3}\partial_{\eta}+a^{-2}\nabla^2
\end{equation}
where $'=\frac{d}{d\eta}$ and $\nabla^2=\partial_{x^i}\partial_{x^i}$. One can check that $\psi(\eta,\vec x)=f(\eta)e^{i\vec k\cdot \vec x}$ is the eigenfunction of $\Delta$ with the corresponding eigenvalue $\lambda$ provided that $f(\eta)$ satisfies
\begin{equation}\label{f}
f''+\frac{a'}{a}f'-(\lambda a^2+\vec k^2)f=0.
\end{equation}
Working up to first order in $\epsilon$ and performing the following substitutions
\begin{align}
f(\eta)=e^{i\omega \eta}(1+\chi(\eta))\\
\lambda=-(\vec k^2+\omega^2)+\delta \lambda
\end{align}
in \eqref{f}, we arrive at
\begin{equation}\label{chi}
\chi''+2i \omega \chi '=\delta \lambda-2(\vec k^2+\omega^2)\epsilon-i \omega \epsilon '.
\end{equation}
Note that to zero'th order in $\epsilon$, $\delta \lambda=\chi=0$. Integrating \eqref{chi} from $\eta=-\infty$ to $\eta=+\infty$ and using \eqref{ep0}, we get $\delta \lambda =0$ even at first order in $\epsilon$. Taking the Fourier transform of \eqref{chi}, we get
\begin{equation}\label{chiF}
\tilde \chi (\Omega)=\frac{2(\vec k^2 +\omega^2)-\omega \Omega}{\Omega(\Omega+2\omega)}\tilde \epsilon(\Omega)
\end{equation}
where Fourier transform defined as $A(\eta)\equiv \int d\Omega~ \tilde A(\Omega) e^{i \Omega \eta}$.

Performing the Fourier transform to get $\chi(\eta)$ back, we need to choose how the contour passes through the poles of $\tilde \chi(\Omega)$. Here, we add a small imaginary number to each term in the denominator of \eqref{chiF},
\begin{equation}
\tilde \chi (\Omega)=\frac{2(\vec k^2 +\omega^2)-\omega \Omega}{(\Omega+i c)(\Omega+2\omega+i c)}\tilde \epsilon(\Omega).
\end{equation}
Note that at the end of calculation $c$ must be taken to zero.
The massless Green's function is given as
\begin{equation}
G(\eta,\vec x;\eta ',\vec x')
=G_0(\eta,\vec x;\eta ',\vec x')+G_1(\eta,\vec x;\eta ',\vec x')\label{FRWGreen}
\end{equation}
where
\begin{align}
G_0(\eta,\vec x;\eta ',\vec x')\equiv \int \frac{d\omega d^2k}{(2\pi)^3}\frac{1}{\omega^2+\vec k^2}e^{i\vec k \cdot(\vec x -\vec x')}e^{i\omega(\eta -\eta ')},\\
G_1(\eta,\vec x;\eta ',\vec x')\equiv \int \frac{d\omega d^2k}{(2\pi)^3}\frac{1}{\omega^2+\vec k^2}e^{i\vec k \cdot(\vec x -\vec x')}e^{i\omega(\eta -\eta ')}\notag\\
\times \left(\chi_{\omega,\vec k}(\eta)+\chi^*_{\omega,\vec k}(\eta')\right).
\end{align}
We first put a cut-off on the eigenvalues of the Laplace operator as follows:
\begin{equation}
\omega^2+\vec k^2 \le \Lambda^2.
\end{equation}
Then, we substitute \eqref{FRWGreen} with the cutoff $\Lambda$ into \eqref{BLmetric}. After some manipulations we arrive at the following
\begin{equation}\label{FRWBLmetric}
g^{\Lambda}_{\alpha \beta}(\eta)=\frac{\pi^2}{36}\delta_{\alpha \beta}+h^{\Lambda}_{\alpha \beta}(\eta)
\end{equation}
where
\begin{equation}\label{h}
h^{\Lambda}_{\alpha \beta}=\frac{1}{(4\pi)^2}\left[\frac{G_{1;\alpha \beta}(\eta)}{G_0^3}-\frac{4\pi^4}{3}\frac{G_1(\eta)}{G_0}\delta_{\alpha \beta}\right]
\end{equation}
and 

\begin{align}
&G_0\equiv G_0(\eta,\vec x;\eta,\vec x)=\frac{4\pi \Lambda}{(2\pi)^3}\\
&G_1(\eta)\equiv G_1(\eta,\vec x;\eta,\vec x)\\
&G_{1;\alpha \beta}(\eta)\equiv \partial_{x^{\alpha}}\partial_{x'^{\beta}}G_1(\eta,\vec x;\eta ',\vec x')|_{x'^{\mu}=x^{\mu}}.
\end{align}
\eqref{FRWBLmetric} shows that the band-limited metric is the flat metric (with the universal prefactor) with additional perturbations. Let us now 
investigate how these perturbations are related to the original metric perturbations $2 \epsilon(\eta)\delta_{\alpha \beta}$. Does one recover the original metric perturbations with the universal prefactor in the limit $\Lambda \to \infty$?

\subsubsection{$h^{\Lambda}_{ij}$ components}

\begin{figure}
\centering
\includegraphics[scale=0.45]{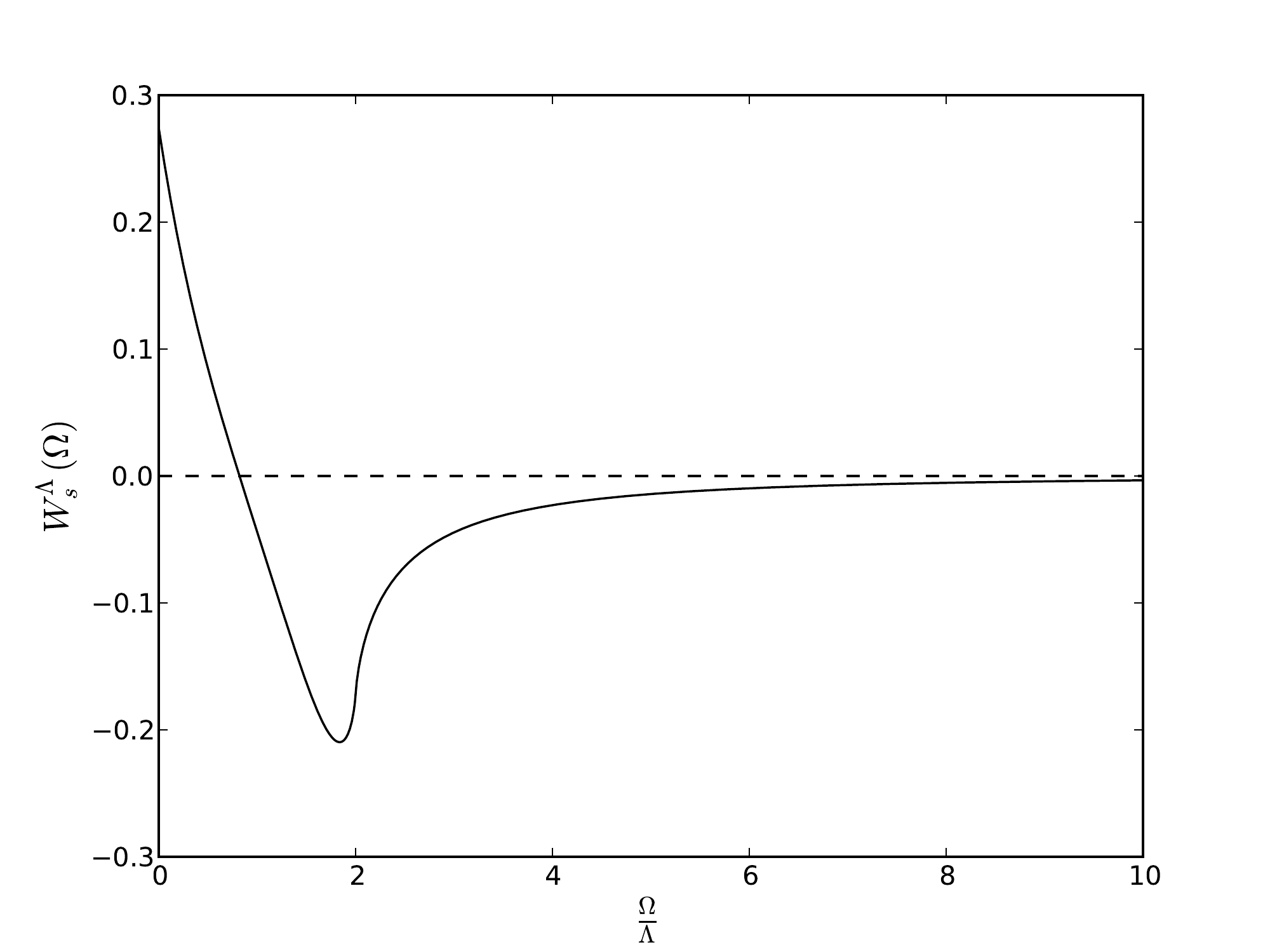}
\caption{High frequency modes compared to the cutoff $\Lambda$ have been damped. As a result, the band-limited metric is becoming more smooth.}
\label{fig:windowS}
\end{figure}

There is no spatio-temporal component to the metric perturbations, since $G_{1;0i}=0$. Spatial components are given by
\begin{align}
h^{\Lambda}_{ij}(\eta)=\frac{\pi}{16}\int^{\Lambda}\frac{d\omega d^2k}{\Lambda^3}\frac{1}{\omega^2+\vec k^2}\left(\chi_{\omega,\vec k}(\eta)+\chi^*_{\omega,\vec k}(\eta)\right)\notag\\
\times \left(k_ik_j-\frac{\Lambda^2}{3}\delta_{ij}\right),
\end{align}
or in Fourier space
\begin{align}
&\tilde h^{\Lambda}_{ij}(\Omega)\notag\\
&=\frac{\pi}{4}\tilde \epsilon(\Omega)\int^{\Lambda}\frac{d\omega d^2k}{\Lambda^3}\frac{1}{\omega^2+\vec k^2}\frac{\vec k^2 +2\omega^2}{\Omega^2-4(\omega+ic)^2}
\left(k_ik_j-\frac{\Lambda^2}{3}\delta_{ij}\right)\notag\\
&=\frac{\pi}{4}\tilde \epsilon(\Omega)\int^{S(1)}d\omega d^2k\frac{1}{\omega^2+\vec k^2}\frac{\vec k^2 +2\omega^2}{\frac{\Omega^2}{\Lambda^2}-4(\omega+ic)^2}
\left(\frac{\vec k^2}{2}-\frac{1}{3}\right)\delta_{ij}
\end{align}
where the last integral is over the region $\omega^2+\vec k^2\le 1$. This means that the spatial part of the original metric perturbation ($2\epsilon(\eta) \delta_{ij}$) in Fourier space is multiplied by the following window function
\begin{equation}
W_s^{\Lambda}(\Omega)=\frac{\pi}{8}\int^{S(1)}d\omega d^2k\frac{1}{\omega^2+\vec k^2}\frac{\vec k^2 +2\omega^2}{\frac{\Omega^2}{\Lambda^2}-4(\omega+ic)^2}
\left(\frac{\vec k^2}{2}-\frac{1}{3}\right)
\end{equation}
Figure \ref{fig:windowS} shows how this window function dampens high frequency modes of metric perturbation and in effect makes the metric more smooth. We can also check that for large values of the UV cutoff, this window function approaches the value $\nu=\frac{\pi^2}{36}$, which is in agreement with our earlier observations. 

\subsubsection{$h^{\Lambda}_{\eta\eta}$ component}
We can use \eqref{h} to find the $\eta\eta$ component of the metric perturbation. Since, $h_{\eta \eta}$ is only a function of $\eta$, however, with a time redefinition we can absorb this term in the definition of time coordinate. So, there is no physical significance to explicitly calculate this metric component. One can directly check that for large values of the cutoff, one obtains
\begin{equation}
h_{\eta \eta}^{\Lambda \to \infty}=\frac{\pi^2}{36}2\epsilon(\eta),
\end{equation}
in agreement with the universality of prefactor $\nu$.


\section{Conclusions and Outlook}

We showed how, in the absence of rods and clocks at sub-atomic scales,  quantum fields can be used, in principle, to measure the curvature of spacetime. Indeed, the imprint that curvature leaves in a scalar propagator, i.e., in the vacuum correlators of a scalar quantum field, suffices to reconstruct the metric and consequently the Riemann tensor. In this sense, the measurement of the Green's function, i.e., of a correlator of quantum fluctuations of fields can replace rods and clocks.

That it should be possible at all to deduce the metric from the propagator was conjectured in \cite{ak-njp} and we here confirm this by giving a constructive method. As a subject for further study, we remark here only that, in \cite{ak-bhamre-prl,ak-njp}, it was also argued that, at least in the case of compact Riemannian manifolds, the mere spectra of the quantum noise on manifolds should suffice to deduce their metric, although in dimensions higher than two, the spectra also of certain tensorial fields should be needed.   

Here, we continued within the framework of euclidean quantum field theory, where we investigated how a hard natural ultraviolet cutoff limits the maximal spatial resolution with which one can reconstruct the metric from the propagator. We found that the metric, expressed terms of the propagator, exhibits characteristic oscillations as the natural UV cutoff scale is approached. These oscillations are generally unobservable in the sense that they should be washed out by the natural ultraviolet cutoff. However, it is conceivable that, through the amplifying effect of cosmic inflation, such oscillations in the metric may have left a signature in the cosmic microwave background. 

To this end, it will be necessary and very interesting to study the covariant natural hard ultraviolet cutoff also in the case of Lorentzian signature. This is nontrivial because, for example, while a hard cutoff makes the Green's function in Euclidean case finite even in the coincidence limit, the corresponding hard cutoff on the Lorentzian Green's function is still divergent in the coincidence limit and on the light cone. This is currently being investigated, see \cite{KempfMartinChatwin}.

Finally, since the tools of Shannon sampling that we applied here to implement an ultraviolet cutoff originate in information theory, it should be very interesting to explore the information-theoretic implications of our findings here. In this context, see e.g., \cite{ak-qg-qc,Lloyd}.




\appendix
\section{Green's function to metric: $D$-dimensional Euclidean space}
\label{app:GtogDEuc}
Here we will prove \eqref{eucGtog} for $D\ge4$.
(Proof for $D=3$ is contained in the main text.)
Let us start with some notation:
\begin{align}
G(x,y)&=f(r_{xy})\\ 
f(r)&=\frac{(2\pi)^{-\frac{D}{2}}}{r^{D-2}}(mr)^{\frac{D}{2}-1}K_{\frac{D}{2}-1}(mr)\\
r_{xy}^2&=\sum_{i=1}^{D}(x^i-y^i)^2.
\end{align}
Let 
\begin{equation}
\tilde{g}_{ij}(x,y)\equiv-\frac{1}{2}\left[\frac{\Gamma(D/2-1)}{4\pi^{D/2}}\right]^{\frac{2}{D-2}}
\frac{\partial}{\partial x^i}\frac{\partial}{\partial y^j}
\left(G(x,y)^{\frac{2}{2-D}}\right).
\end{equation}
Then proving \eqref{eucGtog} is equivalent to showing
\begin{equation}
\lim_{x\to y}\tilde{g}_{ij}(x,y)=\delta_{ij}.
\label{gabxylim}
\end{equation}
It can be checked that
\begin{equation}
\tilde{g}_{ij}(x,y)
=h_1(r_{xy})\delta_{ij}+h_2(r_{xy})
\delta_{ik}\delta_{jl}(x^k-y^k)(x^l-y^l),
\label{noLimEuc}
\end{equation}
where
\begin{align}
h_1(r)&=\frac{1}{2-D}\left[\frac{\Gamma(D/2-1)}{4\pi^{D/2}}\right]^{\frac{2}{D-2}}
r^{-1}f(r)^{\frac{D}{2-D}}f'(r),\\
h_2(r)&=\frac{1}{2-D}\left[\frac{\Gamma(D/2-1)}{4\pi^{D/2}}\right]^{\frac{2}{D-2}}
r^{-2}f(r)^{\frac{2D-2}{2-D}}\notag\\
&\times \left[\frac{D}{2-D}f'(r)^2+f(r)f''(r)-r^{-1}f(r)f'(r)\right].
\end{align}
Since we are interested in the $x\to y$ limit of \eqref{noLimEuc},
it suffices to know the  behaviour of $f(r)$ for small $r$:
(see e.g. 10.31.1, 10.25.2, and 10.27.4 of \cite{bib1})
\begin{align}
f(r)&\xrightarrow[D>4]{mr\to0}\frac{\Gamma(D/2-1)}{4\pi^{D/2}r^{D-2}}
\left[1+\frac{(mr)^2}{2(4-D)}+\cdots\right],\\
f(r)&\xrightarrow[D=4]{mr\to0}\frac{1}{4\pi^2r^2}
\left[1+\frac{(mr)^2}{2}\ln(mr)+\cdots\right].
\label{EucMassLimCorr}
\end{align}
Substituting this back into the definition of
$h_1(r)$ and $h_2(r)$ we find
\begin{align}
h_1(r)&\xrightarrow{mr\to0}1+\cdots\label{h1limEuc}\\
h_2(r)&\xrightarrow[D>4]{mr\to0}\frac{4m^2}{(4-D)(2-D)}+\cdots\\
h_2(r)&\xrightarrow[D=4]{mr\to0}-2m^2\ln(mr)+\cdots \label{h2limEuc}
\end{align}
where $\cdots$ corresponds to sub-leading terms in the expansion. It then follows directly from \eqref{h1limEuc}--\eqref{h2limEuc} that
\begin{equation}
\lim_{x\to y}h_1(r_{xy})=1,\qquad
\lim_{x\to y}h_2(r_{xy})\delta_{ik}\delta_{jl}(x^k-y^k)(x^l-y^l)=0.\label{limEuc}
\end{equation}
Our desired result \eqref{gabxylim} then follows from combining \eqref{limEuc} and \eqref{noLimEuc}.
\section{Green's function to metric: the $D$-sphere}
\label{app:GtogDSphere}
Here we check that \eqref{GtogCurved} is true for
the D-sphere ($D>2$). Let us start by establishing some notation.

The D-sphere is defined as the surface 
\begin{equation}
\delta_{ij}x^ix^j+(x^{D+1})^2=1
\end{equation}
embedded in $D+1$-dimensional Euclidean space
with metric $ds^2=\delta_{ab}dx^adx^b+(dx^{D+1})^2$,
where $a,b=1,\dots,D$. The induced metric on the
$D$-sphere is given by
\begin{align}
ds^2&=g_{ab}dx^adx^b\\
g_{ab}&=\delta_{ab}+\frac{\delta_{ac}\delta_{bd}x^cx^d}{1-x\cdot x},
\end{align}
where $x\cdot y\equiv \delta_{ab}x^ay^b$.
The Green's function on the $D$-sphere is given by
\footnote{
In the embedding $D+1$-dimensional Euclidean space, $Z(x,y)=\delta_{AB}X^AY^B$ ($A,B=1,\dots,D+1$)
where $X$ and $Y$ are the Euclidean coordinates of $x$ and $y$.
}
\begin{align}
G(x,y)&=f(Z(x,y))\\
f(Z)&=\frac{\Gamma(h_+)\Gamma(h_-)}{(4\pi)^{D/2}\Gamma(D/2)}
F\left(h_+,h_-;\frac{D}{2};
\frac{1+Z}{2}\right)\\
Z(x,y)&=x\cdot y+\sqrt{(1-x\cdot x)(1-y\cdot y)}
\end{align}
where $F$ is the hypergeometric function ${}_2F_1$ and 
\begin{equation}
h_{\pm}=\frac{D-1}{2}\pm \nu, \qquad
\nu^2=\frac{(D-1)^2}{4}-m^2.
\end{equation}
Let
\begin{equation}
\tilde{g}_{ab}(x,y)\equiv
-\frac{1}{2}\left[\frac{\Gamma(D/2-1)}{4\pi^{D/2}}\right]^{\frac{2}{D-2}}
\frac{\partial}{\partial x^a}\frac{\partial}{\partial y^b}
\left(G(x,y)^{\frac{2}{2-D}}\right).
\end{equation}
Then confirming \eqref{GtogCurved} is equivalent to
showing
\begin{equation}
\lim_{x\to y}\tilde{g}_{ab}(x,y)=g_{ab}(y).
\label{toshowD}
\end{equation}
It can be shown using straightforward algebra that
\begin{align}
\tilde{g}_{ab}(x,y)&=
h_1(Z(x,y))\frac{\partial Z}{\partial x^a}\frac{\partial Z}{\partial y^b}+
h_2(Z(x,y))\frac{\partial^2 Z}{\partial x^a\partial y^b},\label{gabxy}\\
h_1(Z)&=\frac{1}{D-2}\left[\frac{\Gamma(D/2-1)}{4\pi^{D/2}}\right]^{\frac{2}{D-2}}f(Z)^{\frac{2-2D}{D-2}}\notag\\
&\times \left[\frac{D}{2-D}f'(Z)^2+f(Z)f''(Z)\right],\\
h_2(Z)&=\frac{1}{D-2}\left[\frac{\Gamma(D/2-1)}{4\pi^{D/2}}\right]^{\frac{2}{D-2}}f(Z)^{\frac{D}{2-D}}f'(Z).
\end{align}
Also:
\begin{align}
\frac{\partial Z}{\partial x^a}&=\delta_{ab}
\left(y^b-x^b\sqrt{\frac{1-y\cdot y}{1-x\cdot x}}\right),\label{Zder1}\\
\frac{\partial Z}{\partial y^b}&=\delta_{ba}
\left(x^a-y^a\sqrt{\frac{1-x\cdot x}{1-y\cdot y}}\right),\\
\frac{\partial^2Z}{\partial x^a\partial y^b}&=
\delta_{ab}+\frac{\delta_{ac}\delta_{bd}x^cy^d}{\sqrt{(1-x\cdot x)(1-y\cdot y)}}.\label{Zder2}
\end{align}
Note that in the $x\to y$ limit, $Z\to 1^-$. Therefore, we have to investigate the leading behaviour of $h_1(Z)$ and $h_2(Z)$ when 
$Z\to 1^-$, which in turn depends on the behaviour of $f(Z)$.
It can be shown from asymptotic properties of the hypergeometric function $F$ that
\begin{equation}
f(Z)\xrightarrow{Z\to 1^-}\frac{\Gamma(D/2-1)}{(4\pi)^{D/2}}\left(\frac{1-Z}{2}\right)^{-\frac{D}{2}+1}\left[1+C(Z)+\cdots\right],
\end{equation}
where
\begin{equation}
 C(Z)=
    \begin{cases}
      -\frac{\sqrt{2}\Gamma(1-\sqrt{1-m^2})\Gamma(1+\sqrt{1-m^2})}{\Gamma(1/2-\sqrt{1-m^2})\Gamma(1/2+\sqrt{1-m^2})}\sqrt{1-Z} & \text{if } D=3\\
      \frac{m^2-2}{2}(1-Z)\ln(1-Z) & \text{if } D=4\\
      \frac{D^2-2D-4m^2}{4(D-4)}(1-Z) & \text{if } D> 4.
    \end{cases}
\end{equation}
Plugging this back into the definition of $h_1(Z)$ and 
$h_2(Z)$ we find for $Z\to 1^-$
\begin{align}
 &h_1(Z)\rightarrow
    \begin{cases}
      \frac{-3\Gamma(1-\sqrt{1-m^2})\Gamma(1+\sqrt{1-m^2})}{2\Gamma(1/2-\sqrt{1-m^2})\Gamma(1/2+\sqrt{1-m^2})}\left(\frac{1-Z}{2}\right)^{-1/2}, D=3\\
      (m^2-2)(1-Z)\ln(1-Z),  ~D=4\\
      \frac{D^2-2D-4m^2}{(D-2)(D-4)},~ D> 4.
    \end{cases}\label{hlim1}\\
&h_2(Z)\rightarrow 1+\cdots.\label{hlim2}
\end{align}
It then follows from \eqref{hlim1}, \eqref{hlim2}, and 
\eqref{Zder1}--\eqref{Zder2} that
\begin{align}
\lim_{x\to y}h_1(Z(x,y))\frac{\partial Z}{\partial x^a}\frac{\partial Z}{\partial y^b}&=0\label{lim1}\\
\lim_{x\to y}h_2(Z(x,y))\frac{\partial^2 Z}{\partial x^a\partial y^b}
&=\delta_{ab}+\frac{\delta_{ac}\delta_{bd}y^cy^d}{1-y\cdot y}=g_{ab}(y).
\label{lim2}
\end{align}
Our desired result \eqref{toshowD} then follows from combining 
\eqref{lim1}, \eqref{lim2}, and \eqref{gabxy}.


\section{Green's function to metric: curved manifolds}
\label{app:GtogCurved}
Here we will prove \eqref{GtogCurved} for all curved
manifolds ($D>2$). 
The main idea of the proof is as follows:
due to the presence of the limit $x\to y$
in \eqref{GtogCurved}, it suffices to know
the behaviour of $G(x,y)$ when $x$ and $y$
are arbitrarily close. We will 
derive first order deviations of $G(x,y)$
from flatness in Riemann normal coordinates (RNC)
(for convenience), and
show that corrections due to curvature do not spoil
the Green's function $\to$ metric prescription \eqref{GtogCurved} when the 
limit $x\to y$ is performed.

We start by a brief review of Riemann normal coordinates (RNC),
for the sake of establishing our notation.
\subsubsection{Riemann Normal Coordinates}
Consider a generic coordinate system $\tilde{x}$ on a 
curved manifold. Starting with this coordinate system, we can construct RNC --which we shall denote by $x$-- about point $P$ which has coordinate
$\tilde{y}$ using the transformation
\begin{equation}
\tilde{x}^{i}-\tilde{y}^{i}=x^{i}-\frac{1}{2}\tilde{\Gamma}^{i}_{jk}(\tilde{y})x^{j}x^{k}+\cdots,
\label{RNC_coord}
\end{equation}
where $\tilde{\Gamma}^{i}_{jk}(\tilde{y})$ 
denote the Christoffel symbols at point $P$
in our original coordinate system. By construction, 
the Christoffel symbols and all first derivatives of the metric
vanish at the origin (i.e. $x=0$) of Riemann normal coordinates:
\begin{equation}
\Gamma^{i}_{jk}(0)=0, \qquad
g_{ij,k}(0)=0.
\end{equation}
It can also be shown that the metric in Riemann normal coordinates
takes the form
\begin{equation}
g_{ij}(x)=\tilde{g}_{ij}(\tilde{y})+\frac{1}{3}\tilde{R}_{iklj}(\tilde{y})x^{k}x^{l}+\cdots,
\end{equation}
where $\tilde{R}_{iklj}(\tilde{y})$ are the components of the Riemann tensor in the original coordinate system at point $P$. 
We can always pick the coordinate system $\tilde{x}$ so that at point $P$ the metric is flat: $\tilde{g}_{ij}(\tilde{y})=\delta_{ij}$ for Riemannian manifold (in the following steps replace $\delta_{ij}$ with $\eta_{ij}$ for Lorentzian manifold).
Furthermore, it can be checked using \eqref{RNC_coord} that
$\tilde{R}_{iklj}(\tilde{y})=R_{iklj}(0)$, where $R_{iklj}(0)$ are the components of the Riemann tensor in RNC at point $P$. Therefore:
\begin{equation}
g_{ij}(x)=\delta_{ij}+\frac{1}{3}R_{iklj}x^{k}x^{l}+\cdots,\\
\end{equation}
where for simplicity of notation we have let $R_{iklj}\equiv R_{iklj}(0)$.
Below we list some more useful relations which we will later make use of:
\begin{equation}
g^{ij}(x)=\delta^{ij}+\delta g^{ij}(x),\qquad
\delta g^{ij}(x)=\frac{1}{3}\tensor{R}{^i_k^j_l}x^{k}x^{l}+\cdots
\label{gdetg1}
\end{equation}
as well as
\begin{equation}
\sqrt{g(x)}=1+\delta \sqrt{g(x)},\qquad
\delta \sqrt{g(x)}=-\frac{1}{6}R_{ij}x^{i}x^{j}+\cdots.
\label{gdetg2}
\end{equation}
It is also useful to note that
\begin{equation}
\partial_i\delta g^{ij}(x)=-\frac{1}{3}\tensor{R}{^j_i}x^i+\cdots,~~~
\partial_i\delta \sqrt{g(x)}=-\frac{1}{3}R_{ij}x^j+\cdots.
\label{pargformula}
\end{equation}

\subsubsection{Singularity structure of Green's function}
The Green's function satisfies the equation
\begin{equation}
\left[\frac{\partial}{\partial x^{i}}
\left(\sqrt{g}~g^{ij}
\frac{\partial}{\partial x^{j}}\right)
-m^2\sqrt{g}\right]G(x,y)
=-\delta^{(D)}(x-y).
\label{GF_Eq}
\end{equation}
In Riemannian geometry \eqref{GF_Eq} has a unique solution. However, in Lorentzian geometry there are different solutions to \eqref{GF_Eq}, corresponding to different boundary conditions, and we have to specify which Green's function we are considering. In Lorentzian case, we choose the solution that corresponds to the Feynman propagator. For the purpose of our proof, it suffices that the solution to \eqref{GF_Eq} asymptotes to the flat space Feynman Green's function in the coincidence limit where the effect of curvature is negligible (for more discussion on this see \cite{BirrellDavies}). A large class of states satisfies this condition. For example all the Hadamard states, considered to be physically reasonable states, satisfy this condition (see \cite{paperofHadamard1} and references therein for more details on Hadamard states and their importance.)

We will solve this equation in Riemann normal coordinates
(to first order)
where $y=0$ is the origin. 
To do so, let
\begin{equation}
G(x)=G^E(x)(1+\delta G(x)),
\label{GEdeltaG}
\end{equation}
where we have used the notation
$G(x,0)=G(x)$ and $G^E(x)$ is the massive flat space Green's function which satisfies
\begin{equation}
\left(\delta^{ij}\frac{\partial}{\partial x^{i}}\frac{\partial}{\partial x^{j}}-m^2\right)G^E(x)=-\delta^{(D)}(x).
\label{eucEq}
\end{equation}
It can be checked that to first order \eqref{GF_Eq} becomes 
\begin{align}
&(1+\delta G + \delta\sqrt{g})(\nabla^2G^E-m^2G^E)
\notag\\
&+2\delta^{ij}\partial_{i}\delta G \partial_{j} G^E+G^E \nabla^2\delta G\notag\\
&+\partial_{i}\delta g^{ij}\partial_{j}G^E\notag\\
&+\delta g^{ij}\partial_{i}\partial_{j}G^E
+\delta^{ij}\partial_{i}\delta\sqrt{g}\partial_{j}G^E=-\delta^{(D)}(x),
\label{pertb1}
\end{align}
where $\delta g^{ij}$ and $\delta\sqrt{g}$ are defined in \eqref{gdetg1} --\eqref{gdetg2}.
Using \eqref{eucEq} and \eqref{pargformula}, \eqref{pertb1} reduces to
\begin{align}
&2\delta^{ij}\partial_{i}\delta G \partial_{j} G^E
+G^E \nabla^2\delta G
+\partial_{i}\delta g^{ij}\partial_{j}G^E\notag\\
&+\delta g^{ij}\partial_{i}\partial_{j}G^E
+\delta^{ij}\partial_{i}\delta\sqrt{g}\partial_{j}G^E=0.
\label{pertb1}
\end{align}
Let $\sigma(x)$ denote half the geodesic distance from the origin\footnote{For Lorentzian manifolds we use $(-,+,+,\cdots)$ signature.}:
\begin{equation}
\sigma(x)=\frac{1}{2}\delta_{ij}x^{i}x^{j}.
\end{equation}
Noting that $G^E$ is only a function of $\sigma$ (see \eqref{eucmassiveG} for Riemannian case solution) and using \eqref{gdetg2},
\eqref{pertb1} takes the simpler form:
\begin{equation}
G^E\nabla^2\delta G+2 \frac{dG^E}{d\sigma}x^{i}\partial_{i}\delta G
-\frac{1}{3}\frac{dG^E}{d\sigma}R_{ij}x^{i}x^{j}=0.
\label{pertb2}
\end{equation}
Using the following ansatz for $\delta G$:
\begin{equation}
\delta G(x)=\chi(\sigma(x))+\frac{1}{12}R_{ij}x^{i}x^{j},
\label{deltaG}
\end{equation}
\eqref{pertb2} reduces to the following equation for $\chi$:
\begin{equation}
2G^E\frac{d^2\chi}{d\sigma^2}+(DG^E+4\sigma\frac{dG^E}{d\sigma})\frac{d\chi}{d\sigma}+\frac{R}{6}G^E=0.
\label{chiEq}
\end{equation}
Since we are interested in the small $\sigma$ limit, we can substitute
$G^E(\sigma)$ in \eqref{chiEq} with its $m^2\sigma\to0$ behaviour
\begin{equation}
G^E(\sigma)\xrightarrow{m^2\sigma\to0}
\frac{\Gamma(D/2-1)}{2(2\pi)^{D/2}\sigma^{D/2-1}}.
\end{equation}
In this case \eqref{chiEq} simplifies to
\begin{equation}
2\sigma\frac{d^2\chi}{d\sigma^2}+(4-D)\frac{d\chi}{d\sigma}+\frac{R}{6}=0.
\label{chiODE}
\end{equation}
The general solution of \eqref{chiODE} is
\begin{equation}
\chi=
    \begin{cases}
        \frac{R}{6(D-4)}\sigma+A\sigma^{D/2-1}+B & \text{if } D\neq4\\
        -\frac{R}{12}\sigma\ln(R\sigma) & \text{if } D=4,
    \end{cases}
\end{equation}
where $A$ and $B$ are constants. Requiring $\chi\to0$ 
as $\sigma\to0$ we find $B=0$. 
Therefore, the leading behaviour of $\chi$ for small $\sigma$ is 
\begin{equation}
\chi(\sigma)\xrightarrow{R\sigma\to 0}
    \begin{cases}
      A (R\sigma)^{1/2}+\cdots & \text{if } D=3\\
      -\frac{R}{12}\sigma\ln(R\sigma)+\cdots & \text{if } D=4\\
      \frac{R}{6(D-4)}\sigma+\cdots & \text{if } D> 4.
    \end{cases}
\end{equation}
For $D=3$, $A$ is a constant which depends on global properties of the manifold (e.g. topology). Plugging this result back into \eqref{deltaG}
and then \eqref{GEdeltaG}, and also using the $m^2\sigma\to0$ behaviour of $G^E$
, we find
\begin{equation}
G(x)\xrightarrow[m^2\sigma\to0]{R\sigma\to0}
\frac{\Gamma(D/2-1)}{2(2\pi)^{D/2}\sigma^{D/2-1}}(1+C(x)+\cdots),
\end{equation}
where
\begin{equation}
C(x)=
    \begin{cases}
      A (R\sigma)^{1/2}+(2m^2\sigma)^{1/2} & \text{if } D=3\\
      \frac{m^2\sigma}{2}\ln(m^2\sigma)-\frac{R}{12}\sigma\ln(R\sigma) & \text{if } D=4\\
      \frac{m^2\sigma}{4-D}+\frac{R\sigma}{6(D-4)}+\frac{1}{12}R_{ij}x^ix^j & \text{if } D> 4.
    \end{cases}
\end{equation}

Going back to our original coordinate system 
(using the coordinate transformation
\eqref{RNC_coord})
and computing the RHS of \eqref{GtogCurved}
we find
\begin{equation}
-\frac{1}{2}\left[\frac{\Gamma(D/2-1)}{4\pi^{D/2}}\right]^{\frac{2}{D-2}}
\lim_{\tilde{x}\to \tilde{y}}
\frac{\partial}{\partial \tilde{x}^i}\frac{\partial}{\partial \tilde{y}^j}
\left(G(\tilde{x},\tilde{y})^{\frac{2}{2-D}}\right)
=\delta_{ij}.
\end{equation}
Therefore, we have verified that \eqref{GtogCurved} is true
in the coordinate system $\tilde{x}$ where the metric is chosen to be flat at point $\tilde{y}$.
Since \eqref{GtogCurved} is a tensorial equality,
however, it follows that it is true in all coordinate systems.

\section{Band-limited flat metric}
\label{app:BLflat}
Here, we want to find the band-limited metric associated to band-limited Green's function \eqref{BLGreenflat} of flat space in $D$ dimension. If we perform derivative in \eqref{BLmetric}, we obtain
\begin{align}\label{appD_eq1}
&g^\Lambda_{ij}(x)=-\frac{1}{2}\left[\frac{\Gamma(D/2-1)}{4\pi^{D/2}}\right]^{\frac{2}{D-2}}\notag\\
&\left(\frac{2\partial_{x^i}\partial_{y^j}G_{\Lambda}(x,y)|_{y=x}}{(2-D)G_{\Lambda}(x,x)^{\frac{D}{D-2}}}+\frac{2D\partial_{x^i}G_{\Lambda}(x,y)\partial_{y^j}G_{\Lambda}(x,y)|_{y=x}}{(D-2)^2G_{\Lambda}(x,x)^{\frac{2D-2}{D-2}}}\right)
\end{align}

Here, we calculate each term separately. If we use \eqref{BLGreenflat}, we get
\begin{equation}
\partial_{x^i}\partial_{y^j}G_{\Lambda}(x,y)|_{y=x}=\int^{\Lambda}\frac{d^Dp}{(2\pi)^D}\frac{p_ip_j}{p^2}=\int^{\Lambda}\frac{d^Dp}{(2\pi)^D}\frac{\delta_{ij}}{D}
\end{equation}
where $p_ip_j$ in the integrand is substituted by $\frac{p^2}{D}\delta_{ij}$. Performing the integral, we end up with
\begin{equation}
\partial_{x^i}\partial_{y^j}G_{\Lambda}(x,y)|_{y=x}=\frac{S_{D-1}}{D^2(2\pi)^D}\Lambda^D\delta_{ij},
\end{equation}
where $S_{D-1}=\frac{2\pi^{D/2}}{\Gamma(D/2)}$ is the area of $D-1$ dimensional unit sphere.

Green's function at coincidence point is given by
\begin{equation}
G_{\Lambda}(x,x)=\int^{\Lambda}\frac{p^{D-1}dp~d\Omega_{D-1}}{(2\pi)^D}\frac{1}{p^2}=\frac{S_{D-1}}{(D-2)(2\pi)^D}\Lambda^{D-2}.
\end{equation}
Finally, since first derivative of band-limited Green's function at coincidence point results in an integral over an odd function, the last term in \eqref{appD_eq1} is zero. Substituting these values back in \eqref{appD_eq1}, we get \eqref{BLmetricflat}.

\section{Universal prefactor for 3-sphere}
\label{app:BLS3}
In this section, we show that the band-limited metric of 3-sphere is the original 3-sphere metric up to the universal constant $\nu=\frac{\pi^2}{36}$ when the UV cutoff is taken to infinity. To this end, let us choose a convenient coordinate system for $S^3$. Recall that our method is diffeomorphism invariant and independent of the chosen coordinate system. We choose the toroidal coordinate system $(\chi,\theta, \phi)$, defined as follows
\begin{align}
x^0=\cos(\chi)\cos(\theta)\\
x^1=\sin(\chi)\cos(\phi)\\
x^2=\sin(\chi)\sin(\phi)\\
x^3=\cos(\chi)\sin(\theta)
\end{align}
where $x^{\mu}$ is a Cartesian coordinate of a point at $R^4$ on a unit sphere. The line element on $S^3$ then reads
\begin{equation}
ds^2=d\chi^2+\cos^2(\chi)d\theta^2+\sin^2(\chi)d\phi^2.
\end{equation}
We also need the eigenvalues and normalized eigenfunctions of the Laplacian on $S^3$ \cite{bib2},
\begin{equation}
\Delta T_{k,m_1,m_2}=-k(k+2)T_{k,m_1,m_2}
\end{equation}
where $k\in \{0,1,2,\cdots\}$, $m_1,m_2\in \{-k/2,\cdots,k/2\}$ and
\begin{align}
T_{k,m_1,m_2}(X)=C_{k,m_1,m_2}\left(\cos(\chi)e^{i\theta}\right)^{l}\left(\sin(\chi)e^{i\phi}\right)^{m}\notag\\
P_{k/2-m_2}^{(m,l)}[\cos(2\chi)]
\end{align}
with $l=m_1+m_2$, $m=m_2-m_1$, $C_{k,m_1,m_2}=\frac{\sqrt{k+1}}{\sqrt{2}\pi}\sqrt{\frac{(k/2+m_2)!(k/2-m_2)!}{(k/2+m_1)!(k/2-m_1)!}}$ and $X$ denotes $\{\chi,\theta,\phi\}$ collectively.
For the purpose of our calculations, we only need the following identities
\begin{align}
&\sum_{m_i=-k/2}^{k/2}(m_1+m_2)^2|T_{k,m_1,m_2}|^2=\frac{k(k+1)^2(k+2)}{6\pi^2} \cos^2(\chi),\label{G_relation1}\\
&\sum_{m_i=-k/2}^{k/2}|T_{k,m_1,m_2}|^2=\frac{(k+1)^2}{2\pi^2},\label{G_relation3}\\
&\sum_{m_i=-k/2}^{k/2}(m_1\pm m_2)|T_{k,m_1,m_2}|^2=0.\label{G_relation4}
\end{align}
Then, the band-limited Green's function (with mass $\mu$) is given by
\begin{equation}
G_L(X,Y)=\sum_{k=0}^{L}\sum_{m_i}\frac{1}{k(k+2)+\mu^2}T_{k,m_1,m_2}(X)T^*_{k,m_1,m_2}(Y).
\end{equation}
Let us find the $\theta\theta$ component of band-limited metric. If we substitute $G_L$ in \eqref{BLmetric} and use (\ref{G_relation1}-\ref{G_relation4}), we get the following
\begin{equation}
g^L_{\theta\theta}=\frac{\pi^2}{12}\frac{N}{M^3}\cos^2(\chi).
\end{equation}
where
\begin{align}
N=\sum_{k=0}^{L}\frac{k(k+1)^2(k+2)}{k(k+2)+\mu^2}\notag\\
M=\sum_{k=0}^{L}\frac{(k+1)^2}{k(k+2)+\mu^2}\notag.
\end{align}
For large values of $L$ (cut-off), $N$ diverges as $L^3/3$ , while $M$ diverges as $L$. So, we get
\begin{equation}
g^{L\rightarrow \infty}_{\theta \theta}=\frac{\pi^2}{36}\cos^2(\chi)=\frac{\pi^2}{36}g_{\theta \theta}.
\end{equation}
The same manipulation can be done for the other components of the band-limited metric, which confirms the universality of the prefactor $\nu=\frac{\pi^2}{36}$. Fortunately, however, we do not need to work out the other components of $g^L$. Since the bandlimited metric is rotationally invariant (because the cut-off is a rotationally invariant cut-off), $g^L$ can only be the metric of $S^3$ up to an overall constant. Therefore, one component of $g^L$ already fixes the prefactor.

\end{document}